# Assumptions Underlying Bell's Inequalities

## Angel G. Valdenebro

Lucent Technologies
Valderrobres, 10. 28022 Madrid. Spain
E-mail: avgonzalez@lucent.com

**Abstract**
There are several versions of Bell's inequalities, proved in different contexts, using different sets of assumptions. The discussions of their experimental violation often disregard some required assumptions and use loose formulations of others. The issue, to judge from recent publications, continues to cause misunderstandings. We present a very simple but general proof of Bell's inequalities, identifying explicitly the complete set of assumptions required.

## 1 Introduction

The experimental violation of Bell's inequalities (BI) is a fascinating topic, not only from the theoretical and experimental points of view, but also from the educational point of view. It can be used to illustrate entanglement, one of the pillars of quantum mechanics (QM), by means of an apparent paradox, in a way well suited to undergraduate QM courses. It can be used as well to introduce and motivate the subject of QM interpretation and concepts such as realism, non-locality and hidden variables, even for general audiences (D'Espagnat 1979; Mermin 1981, 1985). In all cases, a good understanding of the assumptions required to proof BI is absolutely mandatory when discussing their experimental violation by QM.

The basic facts are well known, and can be summarised as follows:
(1) Using only simple statistical arguments and common sense assumptions, it is possible to prove that, in a series of measurements on an ensemble of identical systems, the correlations between different observables must obey some constraints called BI.
(2) For certain experimental configurations, QM predicts results that violate BI.
(3) The results of actual experiments closely follow the predictions of QM, establishing the falsity of BI.

J. S. Bell first established the facts (1) and (2) in 1964 (Bell 1964). Brief and recent accounts of fact (3), with references to the latest experimental results, can be found in (Aspect 1999) and (Grangier 2001).

Because of their important implications for our "common sense" (i.e., our conception of the "physical reality"), BI and their experimental violation have been called "one of the profound scientific discoveries of the century" by Aspect (1999) and even "the most profound discovery of science" by Stapp (1975).

Then, what are, exactly, those important implications? It is obvious that at least one of the assumptions made during the proof of BI must be as false as the inequalities are. But the problem starts exactly here. What are the assumptions? Which one of them is false? There are several versions of BI, proved in different contexts, using different sets of assumptions. Some authors consider some of the assumptions so obvious and undeniable valid, that they do not even mention them during their proofs. The generally accepted conclusion "experiments discard local realism" charges the responsibility of BI on just two loosely defined assumptions (locality and realism). Other authors, reluctant to abandon local realism, look for the culprit, or for experimental loopholes, precisely among the assumptions so obviously innocent to others. Despite the hundreds of papers published on the subject since the seminal works of Bell, it is possible to find in recent literature "refutations of BI" (Adenier 2000), or "solutions of the EPR/Bell paradox" (Tartaglia 1998) based on missed, or misinterpreted assumptions.

The whole problem is interestingly entangled with the interpretation of QM. If QM predictions violate BI, any interpretation of QM should be able to explain which one of the assumptions of BI is false and why.

Our goal is to identify and discuss a complete set of assumptions valid to prove BI. The plan is the following. In section 2 we will define the type of system and experiments to which BI apply. In section 3 we will enounce BI and present a very general and straightforward proof based on only one condition: that the correlations can be derived from a joint probability distribution (JPD). But this condition is not at all obvious or justified by common sense. In sections 4, 5, and 6 we will discuss several simple and common sense assumptions that, as is shown in section 7, combined are sufficient to guarantee the existence of the JPD, and consequently the validity of BI.

## 2  Defining the experiment

Let us first consider a system with four observables: *A*, *A'*, *B,* and *B'*. Notice that we have not said "a quantum mechanical system" but just "a system". Regarding the meaning of "observable", we can perform four different experimental procedures on the system, called "measurements", and, as a result of each of the experimental procedures, we obtain a numerical value called "the outcome of the measurement of the corresponding observable", or, simply, "the value of the observable".

Furthermore, we consider an ensemble of identical systems. Let us suppose that on each individual system we can measure simultaneously two observables, one of *A*-type and other of *B*-type, i.e., we can measure the pairs: (*A,B*), (*A,B'*), (*A',B*), and (*A',B'*), but, possibly, not the pairs (*A,A'*) and (*B,B'*). It is also possible, that we cannot repeat the measurements more that once on the same instance. That is, on one individual system we can measure one observable, or a pair of compatible observables, but after that, the system becomes unusable, and to perform more measurements we need new members of the ensemble.

The *A*-type and *B*-type observables are measured on spatially separated subsystems, using independent experimental devices that can be (in principle) arbitrarily far apart from one another.

Let us denote the sets of possible values of the observables *A, A', B,* and *B'* by: $\{a_i\}, \{a'_j\}, \{b_k\}$, and $\{b'_l\}$, respectively, and suppose that the sets are bounded by ±1. We are not giving up much generality in considering discrete sets: were one of the sets continuum, BI would be still valid, just substituting some sums by integrals during the proof. Accepted that the sets are bounded, it is convenient to rescale the observables so that they are bounded by ±1.

Let us suppose that, for any pair of compatible observables the expected value of their product: $\langle AB \rangle$ is a well-defined quantity. Experimentally, this expected value can be calculated by performing a run of measurements of the pair (*A,B*), multiplying the values obtained in each measurement, adding these products, and dividing by the number of measurements. "Well defined" means that several runs of the experiment must produce the same result. Or, more exactly, that when the number of measurements in each run tends to infinity, the value obtained for the expected value tends to a well defined limit (within the experimental error), no matter how the system instances used for the run were selected.

Most presentations of BI do not talk about the "expected value of the product of two observables" but about "the correlation of two observables". Technically the correlation (Pearson's product-moment correlation) of two observables is defined in statistics as

$$c(A,B) \equiv \left\langle \left( \frac{A - \langle A \rangle}{\Delta A} \right) \left( \frac{B - \langle B \rangle}{\Delta B} \right) \right\rangle = \frac{\langle AB \rangle - \langle A \rangle \langle B \rangle}{\sqrt{\langle A^2 \rangle - \langle A \rangle^2} \cdot \sqrt{\langle B^2 \rangle - \langle B \rangle^2}}. \quad (1)$$

When the observables are centred (null expectation value) and normalised to have dispersion equal to one, their correlation and the expected value of their product are the same thing. In the following we will refer to $\langle AB \rangle$, even if not completely rigorous, as the correlation of the observables *A* and *B*.

Notice that we have not made assumptions based on common sense yet, but just defined the class of experiments to which BI are applicable. If we can only identify two observables (not four) on a system, we do not say that the system violates BI, nor that the system challenges our common sense, but just that BI have nothing to do with that type of system. If we cannot calculate the correlations, because different runs of the experiment lead to different and erratic values, the problem surely has to do with an ill-defined system or a bad experimental setup, and not with BI.

## 3   BI from a Joint Probability Distribution

Given a system with the properties defined in the previous section, with the additional assumption that all the correlations can be derived from a JPD for the values of the observables, the following inequalities hold:

$$\left| \langle AB' \rangle + \langle B'A' \rangle + \langle A'B \rangle - \langle BA \rangle \right| \leq 2. \quad (2)$$

An easy way to remember the expression (2) is to write the observables as a list: *(A,B',A',B)*. In fact, four different independent expressions can be proved, one for each of the four possible lists: *(A,B',A',B), (A,B,A',B'), (A',B',A,B),* and *(A',B,A,B')*. Taking into account the meaning of the absolute value, each expression corresponds to two inequalities, what amounts to a total of eight different inequalities, which are part of the family of inequalities generally known as BI.

The JPD is a function $p(a_i, a'_j, b_k, b'_l)$ that has the usual properties for probability distributions:

$$p(a_i, a'_j, b_k, b'_l) \geq 0, \text{ and } \sum_{i,j,k,l} p(a_i, a'_j, b_k, b'_l) = 1. \quad (3)$$

That all the correlations can be derived from the JPD means that

$$<AB> = \sum_{i,k} a_i b_k \left( \sum_{j,l} p(a_i, a'_j, b_k, b'_l) \right),$$

$$<A'B> = \sum_{j,k} a'_j b_k \left( \sum_{i,l} p(a_i, a'_j, b_k, b'_l) \right),$$

$$<AB'> = \sum_{i,l} a_i b'_l \left( \sum_{j,k} p(a_i, a'_j, b_k, b'_l) \right), \quad (4)$$

$$<A'B'> = \sum_{j,l} a'_j b'_l \left( \sum_{i,k} p(a_i, a'_j, b_k, b'_l) \right).$$

The proof of BI (2) is not difficult. We start with

$$\left| a_i (b'_l - b_k) + a'_j (b'_l + b_k) \right| \leq 2, \quad (5)$$

that can be easily seen to be true for any *i, j, k*, and *l*, taking into account that all the quantities $a_i, b_k, a'_j$, and $b'_l$ are bounded by ±1:

$$\begin{aligned}
&\left| a_i (b'_l - b_k) + a'_j (b'_l + b_k) \right| \\
&\leq \left| a_i (b'_l - b_k) \right| + \left| a'_j (b'_l + b_k) \right| \\
&= |a_i| \left| (b'_l - b_k) \right| + |a'_j| \left| (b'_l + b_k) \right| \\
&\leq \left| (b'_l - b_k) \right| + \left| (b'_l + b_k) \right| \leq 2.
\end{aligned} \quad (6)$$

Note that this inequality contains values of observables that could be impossible to measure simultaneously. This is not a problem because we consider that expression just as an algebraic relation involving *possible* values of the observables, not *actual* values obtained in the same measurement.

Averaging a set of values, all of which are bounded by 2, the average must also be bounded by 2, and we get

$$\sum_{i,j,k,l} p(a_i, a'_j, b_k, b'_l) \left| a_i (b'_l - b_k) + a'_j (b'_l + b_k) \right| \leq 2. \quad (7)$$

Using the fact that the absolute value of an average is less than, or equal to the average of the absolute values, it follows

$$\left| \sum_{i,j,k,l} p(a_i, a'_j, b_k, b'_l) \left[ a_i (b'_l - b_k) + a'_j (b'_l + b_k) \right] \right| \leq 2. \quad (8)$$

Finally, operating and using the expressions of the correlations (4), it is straightforward to obtain BI (2), as we wanted to prove.

Summarising, we have just presented a proof of BI based on only one condition: that the correlations can be derived from a joint probability distribution (JPD). But this condition is not at all obvious or justified by common sense. In next sections we will discuss several simple and common sense assumptions that, as is shown in section 7, combined are sufficient to guarantee the existence of the JPD, and consequently the validity of BI.

## 4 Realism assumption

Our first common sense assumption is realism. Realism maintains that *the outcome of a measurement is not created by the measurement, but corresponds to properties possessed by the measured system prior to the measurement*. Closely related to the realist position, and much easier to enounce rigorously, are the *hidden variables (HV) theories*.

### 4.1 HV Assumption
HV theories are characterised by the following principles:
(1) *The properties possessed by the system are represented by HV*;
(2) *By definition, our ensemble of identical systems is in the same macrostate, however the individual systems on the ensemble can be in different microstates, labelled by different values of the HV*;
(3) *The outcome of a measurement on the system is determined by the value of the HV, directly in the so-called deterministic hidden variable theories, or by means of a probability distribution in the stochastic hidden variable theories*.

For example, in the general case of an stochastic HV theory, the probability of obtaining the values $a_i$ and $b_k$ in a measurement of the pair of observables *A* and *B*, conditioned to the measured system being in the microstate characterised by the HV $\lambda$, must be a well defined quantity:

$$p(A=a_i, B=b_k | \lambda), \qquad (9)$$

with the properties required by probability theory:

$$p(A=a_i, B=b_k | \lambda) \geq 0, \text{ and } \sum_{i,k} p(A=a_i, B=b_k | \lambda) = 1. \qquad (10)$$

### 4.2 Reality is single valued
We have already made one tacit assumption above, and, coherent with our goal, we will state it explicitly: *each measurement has only one actual outcome*. It might seem that this is no assumption at all, but a simple consequence of our definition of "measurement", confirmed by our daily perception of the world around us. Nevertheless, there is one interpretation of QM (the many-worlds interpretation) (Everett, 1957) that bases its explanation of the experimental violation of BI on the falsity of this assumption.

## 5 Fair distribution of microstates assumption

Let us assume that *the distribution of microstates in the ensemble depends only on the macrostate (in fact it is what defines the macrostate)*, and denote this (normalised) distribution by $\rho(\lambda)$. Being a normalised distribution requires:

$$\rho(\lambda) \geq 0, \text{ and } \int \rho(\lambda) d\lambda = 1. \qquad (11)$$

We also assume that, *no matter the pair of compatible observables selected, for any sufficiently long run of joint measurements, the microstates on which the measurements are effectively performed are representative of the same general distribution $\rho(\lambda)$*. Using the distribution of microstates (11), and the distribution of probability of outcomes in each microstate (9), we see that if we measure A and B, the probability of obtaining values $a_i$ and $b_k$ is

$$p(A = a_i, B = b_k) = \int d\lambda \rho(\lambda) p(A = a_i, B = b_k | \lambda), \quad (12)$$

and consequently, the correlation for A and B can be calculated in the form

$$\langle AB \rangle = \sum_{i,k} a_i b_k p(A = a_i, B = b_k) = \sum_{i,k} a_i b_k \int d\lambda \rho(\lambda) p(A = a_i, B = b_k | \lambda). \quad (13)$$

Similar expressions exist for the correlations of all the pairs of compatible observables.

This assumption deserves a careful examination, because there are several factors that could threaten its soundness. In next subsections we will identify more than one supposition underlying it.

### 5.1 No selective efficiency (fair sampling)

*If detection efficiency is not perfect*, i.e., if not all the intended measurements can be successfully completed, we must at least assume that *the efficiency does not depend on the microstate of the system*. Otherwise, the measured systems could be unfair representative of the whole ensemble, and the excessive correlations might be possessed only by the detected systems.

This has been called the "detection loophole" and analysed in detail, for example, by Garg and Mermin (1987), who derive a minimum efficiency (*83%*) above which it is safe to say that the experiments (of the Einstein-Podolsky-Rosen type) violate BI.

### 5.2 No conspiracy

*Measurement settings are independent of the microstate of the system*, i.e., *the choice of observables to be measured does not depend on the HV*. Otherwise, it is clear that the instances chosen to measure a certain pair of observables would not, necessarily be fair representatives of the whole ensemble.

For those who regard this assumption to be obvious, denying it means denying the free will of experimenters to choose the measurements they want to perform. The possible implications of the experimental violation of BI for free will have been recognised and considered not only by philosophers, but also by physicists. Asher Peres, in his work "Existence of Free Will as a Problem of Physics" (Peres 1986) concludes: "*free will assumption is, under usual circumstances, an excellent approximation.*"

For radical determinists this assumption is dubious, if not totally false. The same past that determines a system to be on a certain microstate determines the status of the experimental devices, or, why not? the behaviour of the experimenters. But determinism does not forbid the existence of *stochastically* independent events. What we are assuming is that *choices of the experiments (whether to measure A or A', and whether to measure B or B') are performed in a way not correlated with the microstate of the system*.

Note that accepting conspiracy amounts to denying the possibility of experimental knowledge. Were it true, nature could force experimenters to measure what she wants, and when she wants, hiding whatever she does not like physicists to see.

### 5.3 No Backward Causation

*HV are independent of later measurement settings. A posterior event (the choice of experiments) cannot influence a previous fact (the microstate of the system).* Otherwise,

the distribution of microstates would not depend only on the macrostate of the system, but also on future measurements.

One of the main arguments against backward causation is that it could lead to bizarre paradoxes. A classical example involves a person causing the death of his mother before his own birth.

Evident as it may seem, this assumption has been criticised (Costa de Beauregard 1953, 1977; Price 1994, 2001). Let us quote Price when defending his case against causal paradoxes objections:

> "We saw that the causal paradox objection rested on assumption that the claimed earlier effect could be detected in time to prevent the occurrence of its supposed later cause. What does this assumption amount to in the quantum mechanics case? Here the claimed earlier effect is the arrangement of spins in the directions G and H, which are later to be measured. But what would it take to detect this arrangement in any particular case? It would take, clearly, a measurement of the spins of particles concerned in the directions G and H. However, such a measurement is precisely the kind of event which is being claimed to have this earlier effect. So there seems to be no way to set up the experiment whose contradictory results would constitute a causal paradox. By the time the earlier effect has been detected its later cause has already taken place." (Price 1994).

The transactional interpretation of QM, proposed by John G. Cramer (1986) explicitly resorts to backward causation. In exchange it is simultaneously local and realistic.

## 6  Factorability assumption (Locality)

Given the spatial separation of the *A*-type and *B*-type measurements, it seems reasonable to assume that, *the probability of obtaining the value $a_i$ for A, depends only on the microstate of the system, and not on another measurement performed arbitrarily far away, with an independent experimental device*. We will write this probability $p(A = a_i | \lambda)$. Notice that the decision on what observable (*B*, *B'*, or none of them) will be measured simultaneously with *A*, could be taken, in principle, at the last moment, when there is no time for any signal to travel, not even at the speed of light, to the place where *A* is being measured, before the measurement is completed.

In actual experiments, every known source of possible interaction between the experimental devices in "A" and "B" sides must be avoided. However, if the separation is not enough to completely discard the exchange of subluminal signals during measurements, the so-called "locality loophole" or "lightcone loophole" is not completely closed.

Following the laws of conditional probability we can write the probability (9) of measuring the values $a_i$ for *A*, and $b_k$ for *B*, conditioned to the system being in the microstate denoted by $\lambda$, as

$$p(A = a_i, B = b_k | \lambda) = p(A = a_i | \lambda, B = b_k) p(B = b_k | \lambda), \qquad (14)$$

where $p(A = a_i | \lambda, B = b_k)$ represents the probability of measuring the value $a_i$ for $A$ conditioned to the system being in the microstate denoted by $\lambda$ *and* measuring the value $b_k$ for $B$. But our present premise is precisely that

$$p(A = a_i | \lambda, B = b_k) = p(A = a_i | \lambda). \tag{15}$$

Hence, *obtaining the value $a_i$ for A and obtaining the value $b_k$ for B (on a certain microstate of the system) are stochastically independent events, and we can factorise the probability* (9) in the form

$$p(A = a_i, B = b_k | \lambda) = p(A = a_i | \lambda) p(B = b_k | \lambda). \tag{16}$$

This assumption, expressed in any of the equivalent forms (15) or (16), is called Bell-locality condition, factorability, conditional stochastic independence, or "strong locality" (Jarret and Ballentine 1987). Some philosophers prefer a terminology introduced by Hans Reichenbach (1956) in the context of his theory of probabilistic causation, and call this assumption "screening-off". Formally, if $P(X|Y \& Z) = P(X|Y)$, then $Y$ is said to screen $Z$ off from $X$. Intuitively, $Y$ renders $Z$ probabilistically irrelevant to $X$. In our case, being in the microstate $\lambda$ screens the event $B = b_k$ (measuring $B$ and obtaining the outcome $b_k$) off from the event $A = a_i$. And, symmetrically, $\lambda$ screens $A = a_i$ off from $B = b_k$. In other words, *the correlation between A and B is fully explained by the fact that each one of them is correlated to the microstate of the system ( $\lambda$ ).* Factorability prohibits additional correlations between $A$ and $B$ not screened-off by the microstate of the system.

In summary, the premise that we really need to prove BI is factorability, and we have justified this assumption with an argument based on locality considerations. Suppose that we put the blame of the falsity of BI on this premise and decide to reject factorability. Are we required to give up locality completely? Do we need to abolish Einstein's theory of relativity? The answer is that factorability is a stronger assumption than the relativistic locality required by special relativity. We can reject factorability without rejecting relativistic locality.

What is ruled out by Einstein's relativity is any relation between spacelike separated events that requires a unique temporal ordering. Superluminal signals are forbidden by relativity because the event "send a signal" always precedes the event "receive a signal".

Thus, a first requirement imposed by relativity is that no change of experimental parameters at "B" (the decision of measuring $B'$ instead of $B$) can be detected by a change in the probability distribution of the outcomes at "A". Otherwise, the change of the experimental settings at "B" could be used as a kind of superluminal telegraph to send signals to "A". Formally, in the Bell type experiments, relativistic locality imposes the following no-superluminal signalling condition:

$$\int \sum_j p(A = a_i, B = b_j | \lambda) \rho(\lambda) d\lambda$$
$$= \int \sum_k p(A = a_i, B' = b'_k | \lambda) \rho(\lambda) d\lambda \equiv p(A = a_i). \tag{17}$$

Factorability condition (16) is stronger than (17), among other things, because factorability is imposed microstate by microstate while the "no-superluminal signalling" condition is imposed only on the average over the distribution of microstates. It is easy to prove that factorability implies no-superluminal signalling:

$$\int \sum_j p(A = a_i, B = b_j | \lambda) \rho(\lambda) d\lambda$$
$$= \int \sum_j p(A = a_i | \lambda) p(B = b_j | \lambda) \rho(\lambda) d\lambda$$
$$= \int d\lambda \rho(\lambda) p(A = a_i | \lambda) \sum_j p(B = b_j | \lambda) \quad (18)$$
$$= \int d\lambda \rho(\lambda) p(A = a_i | \lambda)$$
$$= p(A = a_i),$$

but no-superluminal signalling does not imply factorability.

To explore deeper into the relation between factorability and relativistic locality we need to introduce some terminology.

## 6.1 Parameter Independence and Outcome Independence

According to Jarrett (1984) the assumption of conditional stochastic independence (factorability) is really a combination of two other assumptions. He calls these premises "simple locality" and "predictive completeness", but we will adopt Shimony's terminology (Shimony 1986) and use "parameter independence" and "outcome independence" instead.

Parameter independence states that, for a given microstate, the probability of an outcome of an observation on the "A" side is (stochastically) independent of the experimental setting (the parameters of the experimental device) on the "B" side. Having in mind that, by definition $p(A = a_i | B, \lambda) = \sum_j p(A = a_i, B = b_j | \lambda)$, parameter independence means

$$p(A = a_i | B, \lambda) = p(A = a_i | B', \lambda)$$
$$= p(A = a_i | \lambda). \quad (19)$$

Outcome independence states that, for a given microstate, the probability of an outcome of an observation on the "A" side is (stochastically) independent of the outcome of the observations on the "B" side. In our notation

$$p(A = a_i | B = b_k, \lambda) = p(A = a_i | B = b_j, \lambda)$$
$$= p(A = a_i | B, \lambda). \quad (20)$$

The name used by Jarret for this condition (predictive completeness) is coherent with the way he enunciates it (Ballentine and Jarret 1987): "A state description is said to be predictively complete with respect to a measurement *A* if *the results* of measurements other than *A* provide no information relevant to predicting the result of *A* that is not already contained in that state description."

In our previous discussion we have combined both premises. Wen we said that something is independent of the event $B = b_k$, we meant that: (a) it is independent of the choice of measuring *B* instead of *B'* (parameter independence), and (b) it is independent of the fact of obtaining precisely the outcome $b_k$ for that measurement (outcome independence). That is, in (15) we have combined (20) and (19). Violation of either or both conditions can account for the violation of factorability.

In general, it is agreed that parameter independence is more fundamental than outcome independence. The observer can control experimental parameters but cannot control outcomes.

It is easy to prove that parameter independence (19) implies the no-superluminal signalling condition (17). Both conditions state the independence of the probabilities of the different outcomes at "A", and the setting of experimental parameters at "B": no-superluminal signalling for the average over the distribution of microstates, and parameter independence for each microstate. Explicitly

$$\int \sum_j p(A=a_i, B=b_j|\lambda)\rho(\lambda)d\lambda$$
$$= \int p(A=a_i|B,\lambda)\rho(\lambda)d\lambda$$
$$= \int p(A=a_i|B',\lambda)\rho(\lambda)d\lambda \quad (21)$$
$$= \int \sum_k p(A=a_i, B'=b'_k|\lambda)\rho(\lambda)d\lambda.$$

Some authors argue that, to assure relativistic locality, no-superluminal signalling is not enough, and parameter independence is also needed. A violation of parameter independence entails that, for a given microstate, the statistical distribution of the outcomes at "A" depends on the instantaneous setting of experimental parameters at "B". And this implies (they say) a relation of distant simultaneity that it is not compatible with relativity. That is why, in my opinion, Jarrett identifies parameter independence with relativistic locality, and calls this condition "simple locality". Even if microstates are not directly observable because observers must always work with runs of measurements and, consequently, with averages over the distribution of microstates, the following question is rather disturbing: in a given microstate, what would be the probability of a certain outcome in a reference frame where the observation at "A" is performed before the experimental parameters at "B" are set?. Whether the previous question is important or not can be discussed, because that probability is not directly observable. Some other authors prefer to stick to the no-superluminal signalling condition as the correct formulation of relativistic locality. What everybody agrees is that parameter independence assures relativistic locality.

On the other side, violation of outcome independence does not conflict with relativity. The correlation between outcomes at "A" and outcomes at "B" it is not meaningful until both measurement have been performed (it does not matter the order), and cannot be detected until observers put together their results (let us say by classical, subluminal telephone) and compare them.

The relations among the different "locality" conditions are summarised in the diagram below

$$\text{factorability} \Leftrightarrow \begin{cases} \text{outcome independence} \\ \\ \text{parameter independence} \overset{?}{\Leftrightarrow} \text{relativistic locality} \end{cases} \quad (22)$$
$$\Downarrow$$
$$\text{no-superluminal signalling} \overset{?}{\Leftrightarrow} \text{relativistic locality}$$

As a conclusion of this section, rejecting factorability forces us to accept the existence of superluminal influences between entangled quantum subsystems, capable of explaining the excess correlations not explained by the common past (the microstate) of the subsystems. We can call this influences "spooky action at a distance" like Einstein, "passion at a distance" like Shimony, or even "fashion at a distance" like Mermin (1999), but we cannot deny them. Peaceful coexistence with relativity is possible if these influences are mutual, of a kind such that there is not a unique way to decide who is "influencing" and who is being "influenced", and such that cannot be used for superluminal signalling. Coexistence with relativity is assured if these influences violate outcome independence but do not violate parameter independence.

## 7  Combining the assumptions

Loosely speaking, HV assumption guarantees us that correlations correspond to objective properties of the system. Fair distribution assumption means that correlations are not enforced by a biased choice of measured systems. Locality prevents correlations from being enforced by action at a distance. Altogether put boundaries to correlations, namely, BI.

But let us prove it formally. We can define

$$p(a_i, a'_j, b_k, b'_l) \equiv \int d\lambda \rho(\lambda) p(A = a_i | \lambda) p(A' = a'_j | \lambda) p(B = b_k | \lambda) p(B' = b'_l | \lambda), \quad (23)$$

were $p(A = a_i | \lambda)$, and the three similar factors for the other observables, take their meaning from the HV assumption, while $\rho(\lambda)$ takes its from the fair distribution assumption.

Using the factorability assumption (16) we can write

$$p(a_i, a'_j, b_k, b'_l) \equiv \int d\lambda \rho(\lambda) p(A = a_i, B = b_k | \lambda) p(A' = a'_j, B' = b'_l | \lambda). \quad (24)$$

As a part of the HV assumption we have stated that $p(A = a_i, B = b_k | \lambda)$ and $p(A' = a'_j, B' = b'_l | \lambda)$ have the properties (10) of probability distributions. Similarly, as a part of the fair distribution assumption, we have stated that $\rho(\lambda)$ has the properties (11) of a density distribution. Combined, these facts guarantee that $p(a_i, a'_j, b_k, b'_l)$ is also a probability distribution (3): it is non-negative and amounts to one when summed over all its indices.

Using the expression (13) for the correlations, it is not difficult to prove that all the correlations can be derived from $p(a_i, a'_j, b_k, b'_l)$ in the form (4) required by a JPD. In conclusion, HV, fair distribution and factorability, allow us to define a JPD, from which all the correlations can be derived. And, as we have already shown, this is enough to prove BI.

## 8  Conclusion

The following assumptions are sufficient to prove BI:
    "Realism":
        1. Hidden Variables (deterministic or stochastic)
        2. Reality is single valued

"Fair distribution"
>3. No selective efficiency
>4. No conspiracy
>5. No backward causation

"Bell-Locality" (Factorability, screening-off condition, conditional stochastic independence):
>6. Parameter independence
>7. Outcome independence

QM predicts the violation of BI in certain experimental configurations. Therefore, every interpretation of QM must consider one of these assumptions, at least, false. Experimental evidence confirms QM predictions. The "paradox" is that all these assumptions seem to be supported by common sense.

# 9 Appendix: Generalised HV models

An immediate consequence of the previous analysis is the so-called "Theorem of Bell": any HV theory that reproduces QM predictions must violate at least one of the other assumptions of BI. Very often the theorem is stated simply as "any HV theory that reproduces QM predictions must be non local". But we have just seen that BI require some other assumptions besides Bell's locality.

The HV model we have used to prove BI is, obviously, of a certain type. There are several recent attempts to defy Bell's theorem by means of more sophisticated HV models. The question is not if there are other types of HV models. It is a fact that there are, and we will mention a couple of them below. The main questions about those generalised HV models are: (a) Are they are able to reproduce the prediction of QM while respecting all the other assumptions of BI? (Otherwise that models do not contribute anything interesting to the debate of Bell's theorem), and (b) Do they represent "realism"? (Remember that our HV assumption was a way of stating rigorously the "realism" postulate. Everybody agrees that if we completely abandon realism Bell's theorem cannot be proved and there is no "paradox" to be solved).

We will briefly mention two generalised HV models that have received some attention recently. Both of them fail to defy Bell's theorem for different reasons.

## 9.1 HV with memory

In the HV model we have used, it is assumed that the HV describing the microstate of the system in the *n* round of measurements would be independent of measurement choices and outcomes of the previous (*n*-1) rounds. That is what allows us to postulate that in any run of the experiment, the microstates corresponding to the different rounds of measurements are representative of the same distribution $\rho(\lambda)$, without any regard to the particular sequence of measurement choices and outcomes in that run. Models which violate this assumption exploit what it is usually called "memory loophole". Memory loophole has been studied in detail in (Barrett et al 2002), concluding that it allows a violation of BI that tends to zero as the number of rounds in the experiment increases. In other words, if the number of rounds in the experiment is large enough we can safely forget about the possibility of "memory" effects.

## 9.2 Hess and Philipp Model

A model due to Hess and Philipp (2001a, 2001b, 2002) using what they call time-like correlated hidden variables has been shown to violate other assumptions (parameter independence, no conspiracy) if it is really to violate BI and reproduce QM predictions (Myrvold 2002, Gill et al 2002).